\documentclass[twocolumn,superscriptaddress,floatfix,preprintnumbers,prd ,nofootinbib,hyperref]{revtex4-2}
\pdfoutput=1
\usepackage[colorlinks=true,breaklinks=true]{hyperref}
\usepackage[normalem]{ulem}
\usepackage[utf8]{inputenc}
\hypersetup{allcolors=[rgb]{0.0 0.0 0.6},linkcolor=[rgb]{0.75 0.05 0.05}}
\usepackage{amsmath,amssymb}
\usepackage{mathtools}
\usepackage{cancel}
\usepackage[dvipsnames]{xcolor}
\usepackage{float}
\usepackage{letltxmacro}
\LetLtxMacro{\oldcite}{\cite}
\renewcommand{\cite}[1]{\mbox{\oldcite{#1}}}
\newcommand{\mnras}{MNRAS}
\newcommand{\apjl}{ApJL}
\newcommand{\actaa}{AcA}

\usepackage{soul}

\allowdisplaybreaks

\bibpunct{[}{]}{,}{n}{}{,}

\hypersetup{
    colorlinks=true,
    citecolor=[rgb]{.1, .7, .6},
    linkcolor=[rgb]{.2, .55, .95},
    filecolor=magenta,
    urlcolor=[rgb]{.1, .7, .6},
}

\begin{document}

\title{Black hole supercolliders}

\author{Andrew Mummery}
\email[]{andrew.mummery@physics.ox.ac.uk}
\affiliation{Rudolf Peierls Centre for Theoretical Physics, University of Oxford, Parks Road, Oxford OX1 3PU, UK}

\author{Joseph Silk}
\email[]{silk@iap.fr}
\affiliation{Institut d’Astrophysique, UMR 7095 CNRS, Sorbonne Universite, 98bis Bd Arago, 75014 Paris, France}
\affiliation{William H. Miller III Department of Physics and Astronomy, The Johns Hopkins University, Baltimore MD 21218, USA}
\affiliation{Beecroft Institute of Particle Astrophysics and Cosmology, Department of Physics, University of Oxford, Oxford OX1 3RH, UK}


\begin{abstract}
We show that collisions between particles free falling from infinity and a disk of material  plunging off the retrograde innermost stable circular orbit of a near-extremal Kerr black hole is the unique astronomically natural way in which to create a gravitational particle accelerator with center of mass energies at the $10$'s to $100$'s of teraelectronvolt range, in other words a supercollider. 
\end{abstract}

\maketitle

{\it Introduction –} The accretion of material onto an astronomical black hole modifies the properties of the hole, and in particular can act to spin up the hole to (near) maximal rotation on timescales which are short (astronomically speaking). A Schwarzschild black hole with mass parameter $M_\bullet = M_i$ is spun up to maximal rotation as the mass parameter grows to $M_\bullet = \sqrt{6}M_i$  by accretion through a thin disc \cite{Bardeen1970}. At an Eddington-limited rate of accretion the characteristic timescale for this mass growth is the so-called Salpeter timescale $\tau_s \simeq 10^7$ years \cite{Salpeter64}. The capture of some of the radiation field emitted during the accretion process limits this final spin growth \cite{Thorne1974}, although the value of this limiting spin is dependent on the details of the accretion model \cite{Abramowicz80} and when models are allowed to explore parameter values above this classical limit observations have been found to require even higher spins at strong statistical significance \cite{Zhao21}. 

As at least some of the supermassive black holes (SMBHs) which reside in galactic centers are likely to have grown their final $(\sqrt{6} - 1)$ fraction of mass by coherent accretion, and it is reasonable therefore to expect that at least some supermassive black holes will be near maximally rotating. Indeed this is what observations suggest (e.g., \cite{Reynolds13}), a variety of different techniques all suggest that a significant fraction of the SMBH population are rapidly rotating \cite{Reynolds19, Piotrovich22}, a result which appears to hold even at larger redshifts $z\sim 2$ \cite{Trakhtenbrot14}. Similarly, the stellar mass black holes in Galactic X-ray binaries are near-universally found to be near-maximally rotating \cite{Zhao21, Draghis2023}. 

The supermassive black holes in galactic nuclei go through phases of activity and inactivity, simply reflecting the time dependence of the availability of material to accrete. Consider an inactive black hole which was spun up to $a_\bullet = 1- \delta, (\delta\ll 1),$ (we use units in which $GM_\bullet = 1 = c$ throughout) during a previous phase of accretion, which switches on again, perhaps due to the tidal disruption of a star \cite{Rees88}, or due to a new phase of feeding from interstellar gas. On the large radial scales on which this material is fed towards the black hole, no imprint of the spin parameter of the black hole will be felt. This means it is equally likely that this new material will approach the black hole in a prograde or retrograde fashion (formally this equal probability of misalignment only holds on timescales shorter than the Salpeter time for disks with larger angular momenta than their black hole, and at all times for disks with angular momenta significantly smaller than the hole \cite{King05}). 

Accretion of material (when fed at high rates towards the black hole) is mediated through a disk, where the fluid follows approximately circular orbits down to the innermost stable circular orbit, where-after circular orbits become unstable to inward perturbations, and the fluid elements plunge towards the black hole horizon. It is significantly more difficult for a fluid element to counter-rotate against the black hole spin when the flow reaches near-horizon scales, a result which is reflected in the location of this innermost stable circular orbit (ISCO) for retrograde motion $r_{I, r} = 9 - 45\delta/16$, which is significantly larger than that of prograde motion $r_{I, p} = 1 + (4\delta)^{1/3}$ \cite{Bardeen72} (Boyer-Lindquist coordinates are used throughout and these radii are valid only to leading order in $\delta$). The horizon for near-extremal spins is located at $r_+ = 1 + (2\delta)^{1/2}$. This means that material that plunges off the inner regions of a newly-formed retrograde accretion flow of a near extremal Kerr black hole has a large ``run-up'' towards the horizon, during which they undergo a gravitationally dominated acceleration. We shall show in this {\it Letter}  that by the time they reach near horizon scales, if they were to impact upon a particle free-falling from infinity (for example), then the center of mass energies of these collisions can be large $E_{\rm CM} \gtrsim 100 m_0$, where $m_0$ is the rest mass of the particles. In fact these energies scale as $E_{\rm CM} \sim \delta^{-1/4}$, and diverge for extremal spins (a common \cite{Jacobson10}, but not universal \cite{Harada14} divergent scaling). This is an example of the Ba\~nados-Silk-West (BSW) effect \cite{Piran75, BSW2009} for extremal spins and collisions at the horizon, first proposed as a possible route towards detecting hypothetical weakly interacting dark matter particles. 

In recent years the BSW effect has generated a huge amount of interest in the community, with much work demonstrating that collisions resulting in formally arbitrarily high energies are a generic result of near-horizon dynamics of extremal black hole spacetimes (e.g., \cite{Harada14}), with the large proper volume available for ergospheric collisions allowing highly relativistic collision debris to escape to infinity with potentially observable consequences \cite{schnittman2015}. A major remaining problem for relating the BSW effect to observational reality is in providing a robust scenario for loading the ergosphere of a rapidly spinning black hole with appropriate orbits for collisions of multiple particles (i.e., an astronomically reliable feeding scenario for the collider).  

It is the purpose of this {\it Letter} to demonstrate that there is likely only one astronomically natural way to generate a large flux of particles into the ergosphere on orbits which are highly relativistic, and which if they collide with particles will produce large center-of-mass energies. This is an accretion flow plunging off the retrograde ISCO, which for moderate accretion rates will produce a particle flux of $\dot N \sim 10^{44} (M_\bullet/M_\odot) (\mu/100 \, {\rm GeV})^{-1}\, {\rm s}^{-1}$, where $\mu$ is the typical rest-energy scale of the particles in the flow.  For typical astrophysically relevant atomic species this means that center of mass energies for collisions between disk and free-falling particles can hit the $\sim 1-100$'s of TeV range.  It is therefore this particular setup which should be focus of detailed models of the spectrum and flux of outgoing collision remnants. 

{\it Analysis –} The total energy in the center of mass frame resulting from a collision between two particles with rest masses $m_1$ and $m_2$ is 
\begin{equation}
    E_{\rm CM}^2 = m_1^2 + m_2^2 - 2m_1m_2 g_{\mu\nu} u^\mu_{(1)} u^\nu_{(2)}, 
\end{equation}
where we use the mostly positive metric signature. The spacetime metric is $g_{\mu\nu}$ and $u^\mu_{(i)}$ represents the four velocity of the $i^{\rm th}$ particle. This is a simple result from special relativity which holds in general relativity owing to the equivalence principle. 
\begin{figure}
    \includegraphics[width=1.05\linewidth]{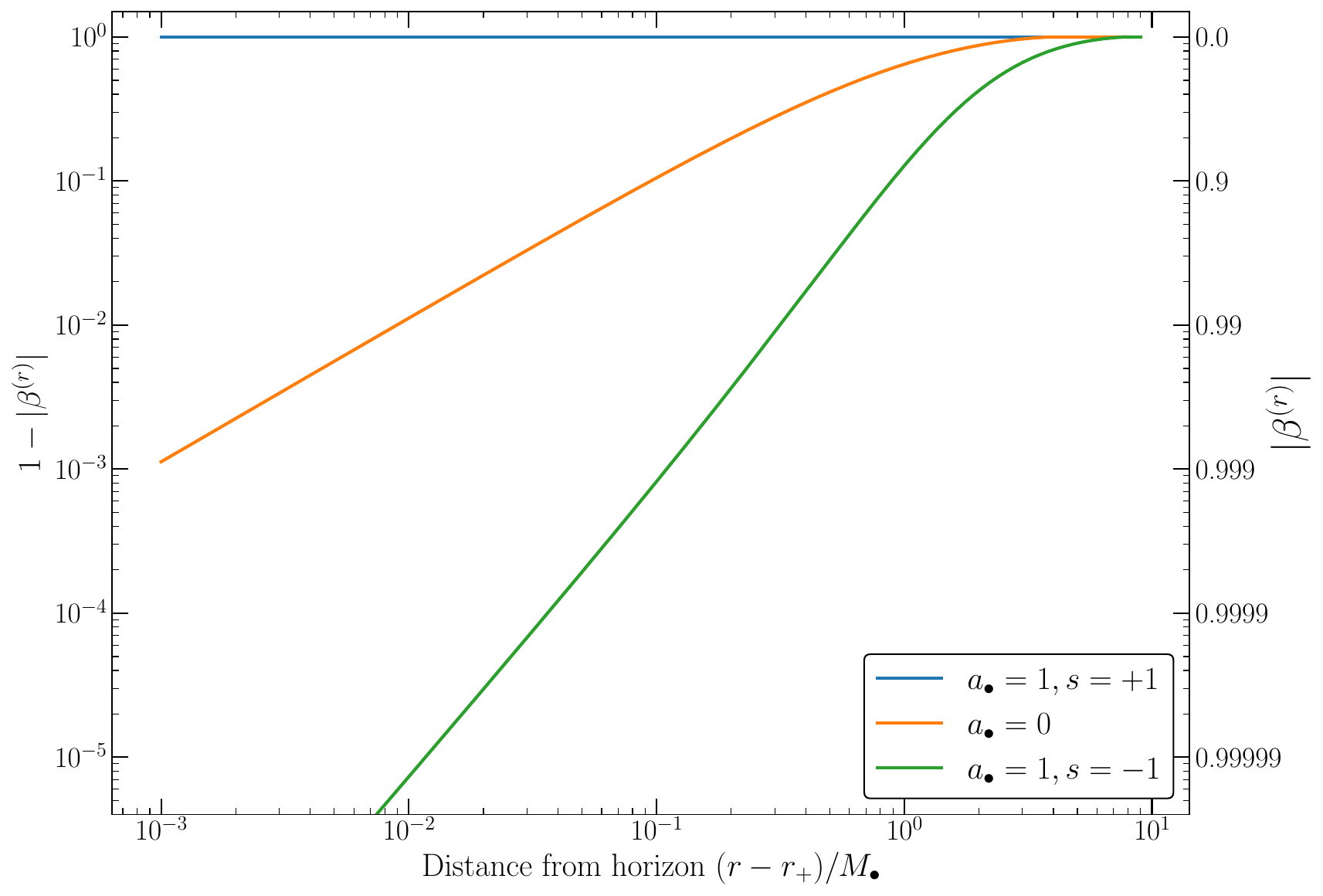}
    \includegraphics[width=.92\linewidth]{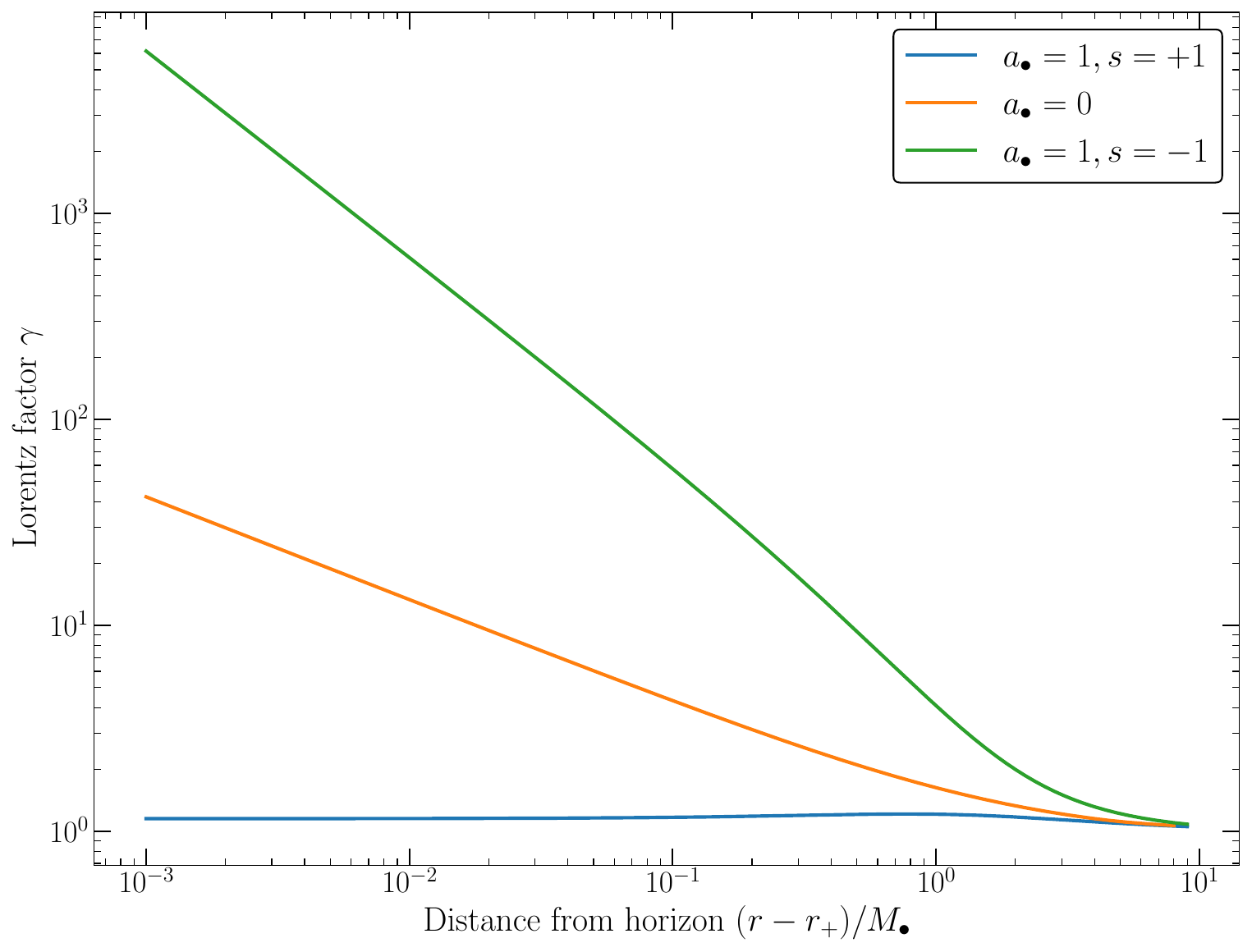}
    \caption{The relativistic Lorentz factor $\gamma$ (lower panel) and the absolute value of the radial velocity $|\beta^{(r)}|$ (upper panel) measured by a zero angular momentum observer for particles spiraling through an accretion flow, around different black hole spins, as a function of the horizon distance. Negative values of $s$ mean that the disk rotates counter to the black hole's spin axis. The larger gap between the ISCO and event horizon for retrograde spins maximizes the relativistic nature of the near-horizon dynamics. }
    \label{fig:dynamics}
\end{figure}

In our picture, one of the four-velocities of the particles is given by that of a particle which has inspiralled from the retrograde innermost stable circular orbit  towards the horizon. The general solution for particle motion in this limit is given by \cite{Cunningham75, MummeryBalbus22PRL}
\begin{equation}
    u^r = - \sqrt{2\over 3r_I} \left({r_I\over r} - 1\right)^{3/2}, 
\end{equation}
with constants of motion $u_\phi = 2\sqrt{3} s(1 - 2 a_\bullet s/3r_I^{1/2})$, and $u_0 = -(1-2/3r_I)^{1/2}$. In these, and future, expressions $0 \leq a_\bullet \leq 1$, and the quantity $s = \hat J_{\rm d} \cdot \hat J_{\rm \bullet} = \pm 1$ represents the relative orientation of the unit vectors of the disk angular momenta ($\hat J_d$) and black hole angular momenta ($\hat J_\bullet$), and we remind the reader that the ISCO radius is itself a function of this orientation.  

To examine the effects of the long ``run-up'' to the horizon for near-extremal retrograde spins, it is illustrative to compute the four-velocity which would be observed by a zero angular momentum observer \cite{Bardeen72}, for whom the spacetime is locally flat. This can be computed in the standard manner 
\begin{equation}
    v^{(a)} = E^{(a)}_{\mu} u^\mu \equiv \gamma\, (1, \vec\beta). 
\end{equation}
We refer the reader to \cite{Bardeen72} for the functional form of the transformation matrix $E^{(a)}_\mu$. We display the radial component of the observed four velocity $\beta^{(r)} \equiv v^{(r)}/v^{(0)}$ (we use bracketed indices to distinguish locally measured quantities from truly covariant quantities), and the locally observed Lorentz factor of the plunging particles $\gamma \equiv v^{(0)}$ in Fig. \ref{fig:dynamics}. Outside of the ISCO we assume the particles follow circular orbits. 

In Fig. \ref{fig:dynamics}, it is clear to see that the optimum setup for maximizing the relativistic nature of near-horizon collisions is to orientate the disk in a retrograde fashion with respect to the black hole spin axis, and to seek a maximal black hole spin parameter. Fortunately, as we discussed above, this is a relatively natural astronomical state of affairs.

\begin{figure}
    \centering
    \includegraphics[width=1\linewidth]{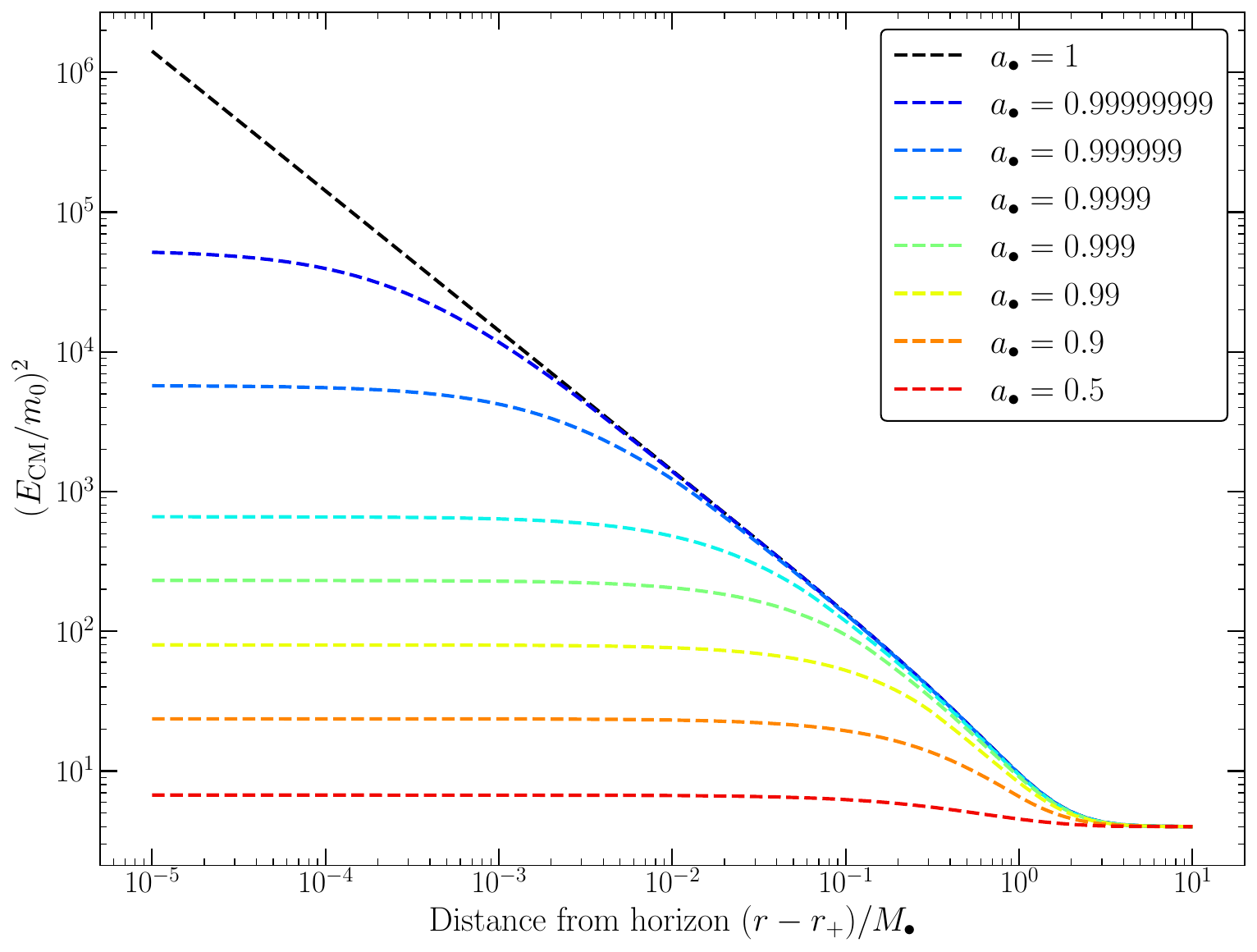}
    \caption{The center of mass energies, as a function of the distance from the horizon at which the collision occurred, of a collision between a prograde ($s=+1$) and retrograde ($s=-1$) inspiralling particle, for different black hole spins. We assume $m_1 = m_2 \equiv m_0$. For an extremal spin the center of mass energy diverges as $E_{\rm CM} \sim 1/(r_c-r_+)^{1/2}$, where $r_c$ is the collision radius, while for near extremal spins this energy plateaus at a scale $E_{\rm CM} \sim \delta^{-1/4}$ as $r_c \to r_+$.  }
    \label{fig:pro+ret}
\end{figure}

Particles in an accretion flow will naturally have some distribution of angular momenta and energies, owing to the turbulent nature of the flow itself. The absolute maximum energy available in a collision within the flow is if one particle had its motion entirely reversed, and was co-rotating with the black hole (with everything else counter-rotating). This is of course an extremely unlikely setup, but it is simple to compute analytically (as has been noted previously \cite{Harada14}), and produces results which are qualitatively similar to the more natural setup of a collision between a disk particles and those particles dropped from infinity, which we shall consider next. 

The center of mass energies, as a function of the distance from the horizon at which the collision occurred, of a collision between a prograde and retrograde inspiralling particle are shown in Fig. \ref{fig:pro+ret}. The color of the line denotes different black hole spins. For an extremal spin, the center of mass energy diverges as $E_{\rm CM} \sim 1/(r_c-r_+)^{1/2}$, where $r_c$ is the collision radius. This can be shown by considering the properties of the relevant contraction  
\begin{multline}    
    g_{\mu \nu} u^\mu_{(1)} u^\nu_{(2)} = \bigg[(1 - 2/r) J_p J_r - (r^2 + a_\bullet^2 + 2 a_\bullet^2/r) E_p E_r \\ + (2a_\bullet/r) (J_p E_r + J_r E_p) + r^2 U^r_p U^r_r\bigg] {1\over \Delta}, 
\end{multline}
where $\Delta \equiv (r-r_+)(r-r_-), r_\pm = 1 \pm \sqrt{1-a_\bullet^2}$. In this expression $J = u_\phi, E = -u_0$ and subscripts $p/r$ represent prograde and retrograde, respectively. It can be shown that while $1/\Delta(r_+, a_\bullet)$ diverges, the numerator of this expression also vanishes for any non-extremal spin parameter, leading to finite results.  

It is interesting however to consider the limit $r\to r_+$ for $a_\bullet = 1-\delta$.  In the near-extremal limit the relevant physical quantities, to lowest order in $\delta$, are 
\begin{align}
    r_{\pm} &= 1 \pm (2\delta)^{1/2} \nonumber, \\
    J_p &= \sqrt{4\over 3} \, (1 + (4\delta)^{1/3}) \nonumber, \\ 
    E_p &= 3^{-1/2} (1 + (4\delta)^{1/3}), \nonumber\\
    U^r_p &= -2\, (6  \delta)^{1/2} / 3 ,
\end{align}
while the retrograde angular momentum $J_r = -22\sqrt{3}/9$, energy $E_r = 5\sqrt{3}/9$,  and radial velocity $U^r_r = -32\sqrt{3}/9$ all have corrections at higher powers of $\delta$. The vanishing factor in the numerator is independent of $\delta$, and is a linear function of the horizon distance $r_c-r_+$. This means that for near extremal spins the energy plateaus at a scale $E_{\rm CM} \sim \delta^{-1/4}$, and diverges as $1/(r_c-r_+)^{1/2}$ for extremal spins owing to the double root of $\Delta$ in this limit. This is an example of the BSW effect.

\begin{figure}
    \centering
    \includegraphics[width=\linewidth]{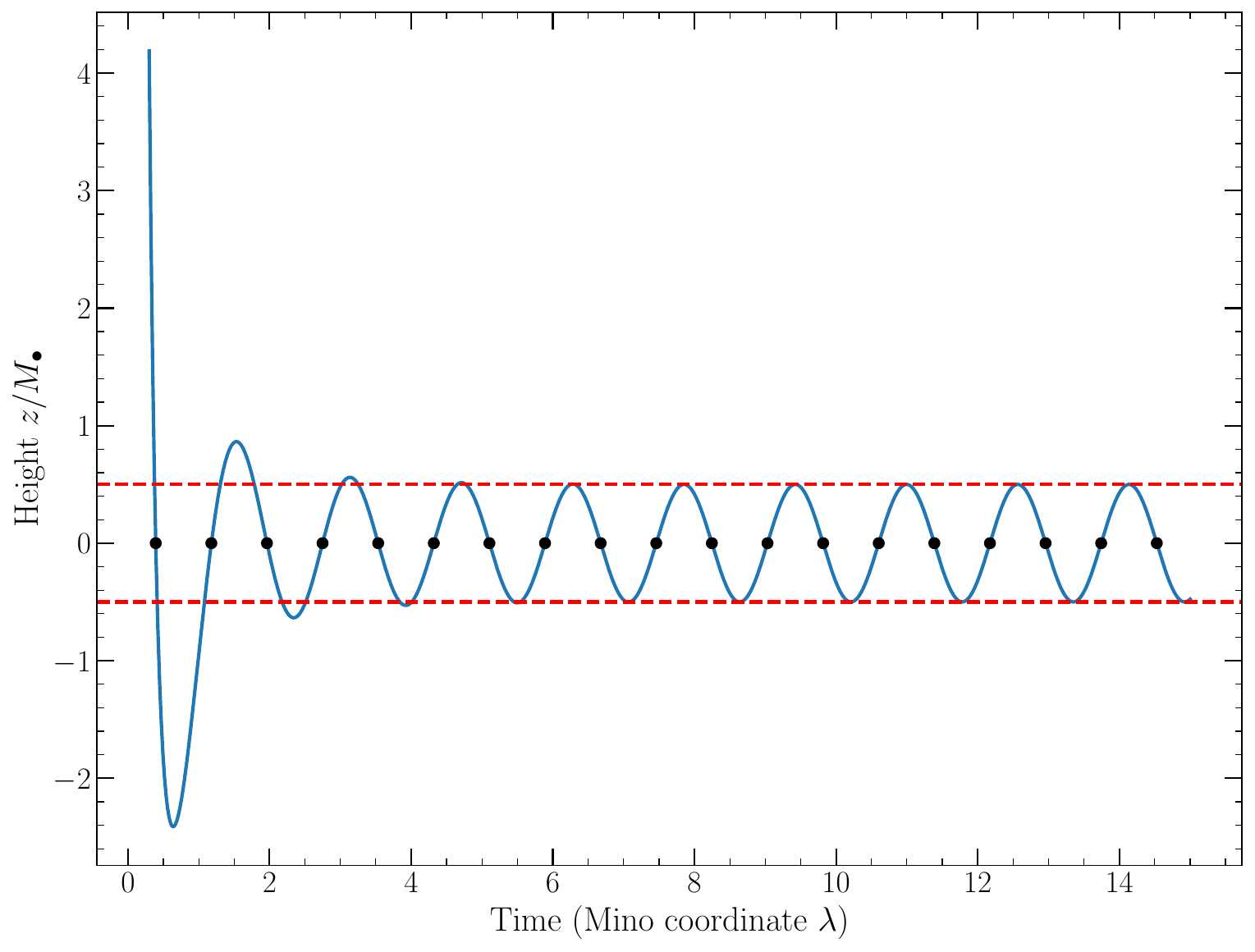}
    \includegraphics[width=\linewidth]{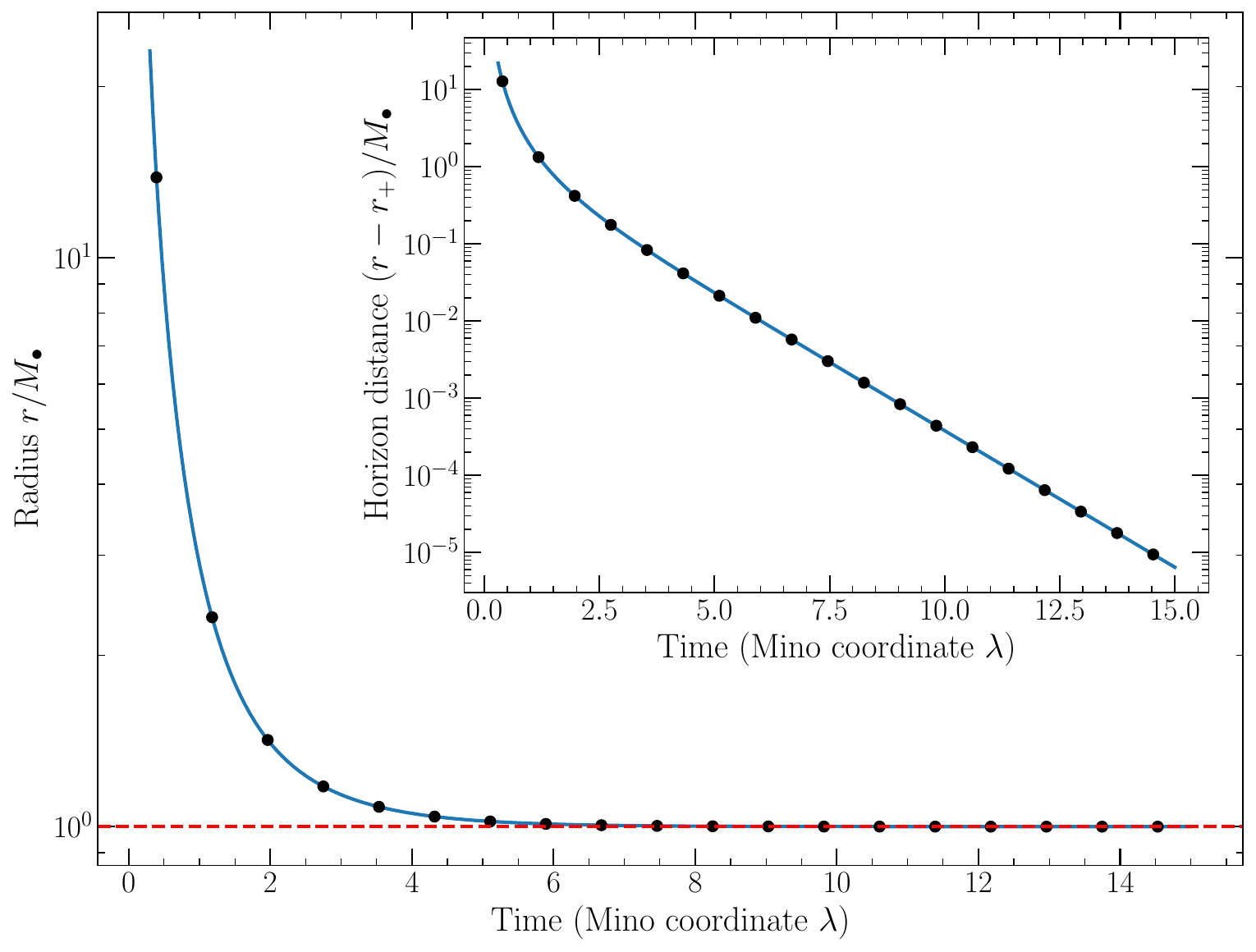}
    \includegraphics[width=\linewidth]{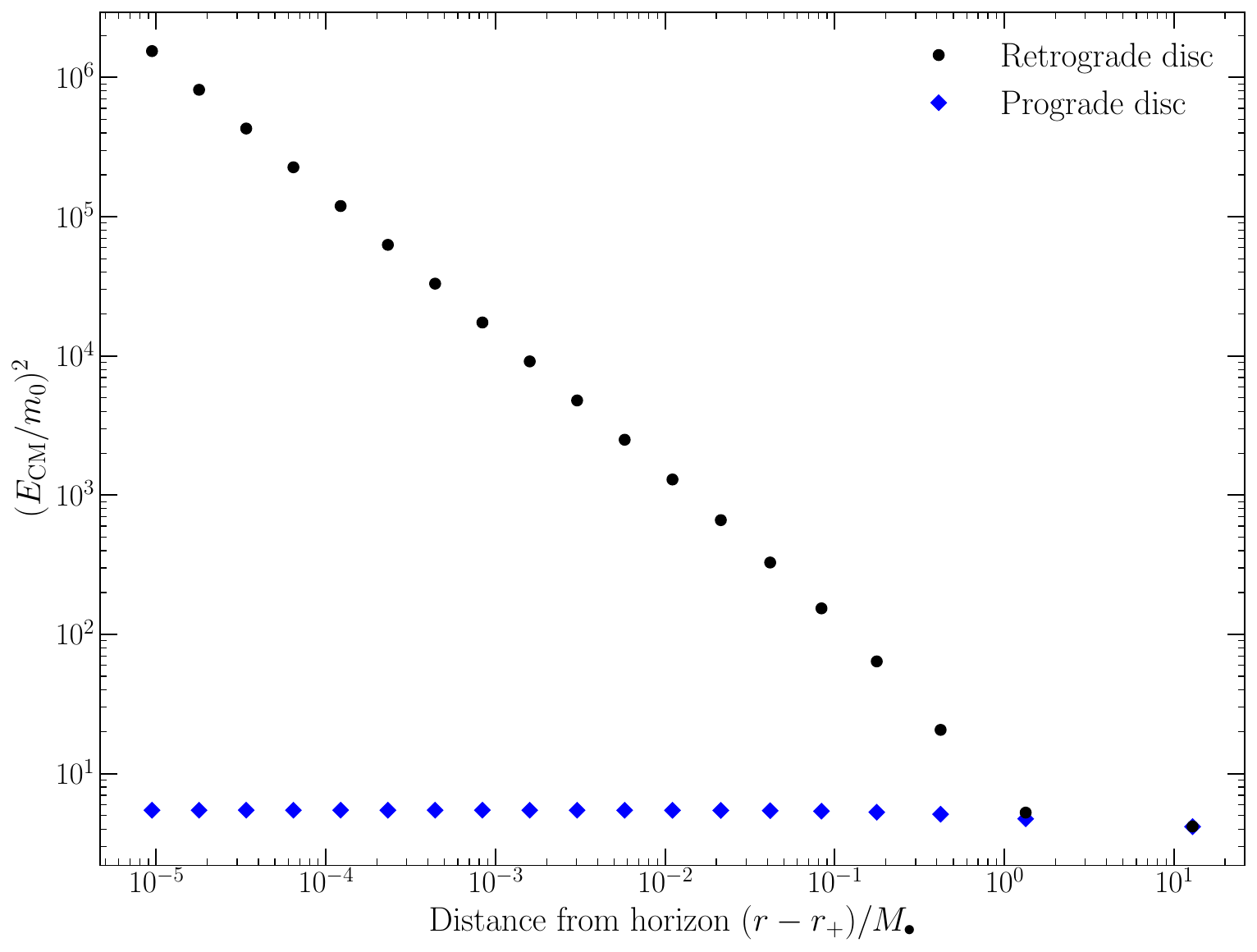}
    \caption{Collisions between a particle free-falling from infinity on an IBSO orbit and a plunging disk particle. Upper: the vertical $z = r\cos\theta$ coordinate of the free-falling particle, which crosses the equatorial plane an infinite number of times (black dots) as it asymptotes towards $r = \chi = r_+ =1$ (red dashed line, middle panel). Both are plotted against the so-called Mino time parameter (see text). Each time the particle crosses the equatorial plane we compute the center of mass energy a collision with a disk particle would have (lower panel, assuming $m_1 = m_2 = m_0$, prograde in blue, retrograde in black). As before, the retrograde disk energy diverges as $E_{\rm CM} \sim 1/(r_c-r_+)^{1/2}$, while for near extremal spins this energy plateaus at a scale $E_{\rm CM} \sim \delta^{-1/4}$ as $r_c \to r_+$.  }
    \label{fig:ibso}
\end{figure}

A more realistic set up for a collision is between a particle in the plunging disk and a particle which has fallen in from infinity, having started at rest with respect to the black hole. Again there is a simple analytical system with which to probe the maximal center of mass energies here, a collision between a particle on the innermost bound spherical orbit (IBSO) trajectory and a retrograde plunging particle. The IBSO orbit corresponds to a particle starting at rest at infinity $(E=-u_0=1)$, with precisely the right angular momentum to asymptote towards  a spherical (in the case of an inclined orbit in the Kerr metric) or circular (for equatorial orbits) orbit. The symmetries of this setup allow for a simple analytical treatment, while retaining key features of more general orbits falling from infinity. 

The equations of motion for a test particle infalling with unit energy in the Kerr metric can be expressed in the following manner \citep[see e.g.,][]{Mummery24} 
\begin{align}
    & \left({{\rm d} r \over {\rm d}\lambda}\right)^2 = 2(r-\chi)^2(r-r_2) ,\nonumber\\
    & \left({{\rm d} \theta \over {\rm d}\lambda}\right)^2 = q - l_z^2 \cot^2\theta , 
\end{align}
where we make use of the so-called Mino time parameter $(r^2 + a_\bullet^2 \cos^2\theta){\rm d}\lambda = {\rm d}\tau$ \cite{Mino03}, and the particle will asymptote towards the IBSO radius $\chi$.  The constants of motion are $l_z$ (the axial angular momentum) and $q$ the Carter constant \cite{Carter68}. It will prove more natural to parameterize the orbit by $\chi$ and the asymptotic angle which the velocity vector of the particle makes with the black hole's spin axis $\sin \psi \equiv l_z / \sqrt{l_z^2 + q}$. In this case the orbital motion can be solved analytically (the equation of motion for the variable $h = \cos\theta$ is in fact that of simple harmonic motion with energy $q$)
\begin{align}
    r(\lambda) &= r_2 + (\chi-r_2) \coth^2\left({\lambda}\sqrt{\chi - r_2\over 2}\right), \nonumber \\
    \theta(\lambda) &= \cos^{-1} \left[\cos(\psi) \, \cos\left({\lambda l_z \over \cos \psi}\right)\right]  .
\end{align}
The second radial root is given by $r_2 = a_\bullet^2 l_z^2 \cot^2\psi / (2\chi^2)$, while $l_z = 2\sin\psi\sqrt{\chi^3/(\chi^2-a_\bullet^2\cos^2\psi)}$. The IBSO radius itself is the root of 
\begin{multline}
    \chi^4 - 4\chi^3 - a_\bullet^2(1 - 3 \cos^2\psi)\chi^2 + a_\bullet^4\cos^2\psi \\ + 4a_\bullet \sin \psi \sqrt{\chi^5 - a_\bullet^2\chi^3\cos^2\psi} = 0 .
\end{multline}
We plot the properties of these solutions in Fig. \ref{fig:ibso}, for $\psi = \pi/3$ and $a_\bullet = 1$. The particle falls from infinity, and asymptotes towards $\chi$, which for this set of parameters equals the event horizon $\chi = r_+ = 1$ \cite{Hod13}. The particle crosses the equatorial plane a formally infinite number of times on its trajectory. Each time the particle crosses the equatorial plane we compute the center of mass energy a collision with a plunging disk particle would have, which is shown in the lowest panel. A plunge of a retrograde ISCO is a dramatically more efficient particle accelerator than prograde orbits.  In an identical fashion to the prograde-retrograde collision considered previously, the center of mass energy for a retrograde plunge collision diverges as $E_{\rm CM} \sim 1/(r_c-r_+)^{1/2}$ for this orbit. All IBSO orbits for asymptotic inclination angles $\sin \psi \geq \sqrt{2/3}$ asymptote to $\chi=1$ \cite{Hod13}, and so a reasonable portion of phase space is available to these most energetic collisions. Again, if a finite near-extremity parameter $\delta$ were assumed, then a plateau at $E_{\rm CM} \sim \delta^{-1/4}$ is found. 

As the formation of retrograde accretion flows around near-extremal Kerr supermassive black holes are astronomically rather natural to expect, and as these accretion flows will contain all astrophysically relevant atomic species, the chance of a collision between relativistically plunging material and free-falling particles at all radii down to the horizon may well result in rest-frame energies at the level of 1 to 100's of TeV (or more). The large proper volume available for ergospheric collisions allows highly relativistic collision debris to escape, and we believe that neutrino signatures may be of particular interest.  

{\it Observable consequences -}
The next generation particle supercollider, the Future Circular Collider, is  currently under  study and has a design and construction time-scale  of at least thirty years \cite{benedikt2022}. It is intended to reach particle collision energies in the center-of-mass frame up to   100 TeV and defines the future of high energy particle physics.
We  have shown that plunging accretion flows within the ISCO region around extremely rapidly rotating supermassive black holes, often considered to be ubiquitous in massive galaxy nuclei, can generate particle collision energies up to, and possibly even beyond, this order via feeding the outer ergospheric region.

Speculatively, we can look to otherwise hard-to-explain properties of  energetic (astronomical) particles as a possible example of the effect discussed in this {\it Letter}.   A recent neutrino event with energy hundreds of times larger than the energies previously measured by state-of-the-art detectors such as ICECUBE, which may therefore require such a supermassive black hole supercollider to be produced, is the recent event KM3-2302A at 220 PeV measured by KM3Net \cite{km3netcollaboration2025ultrahighenergyeventkm3230213aglobal}. 

While the issue of whether  SMBH are indeed rapidly rotating is currently under debate (e.g., \cite{Zhao21}), we note that the plausible intervention of a nearby IMBH can further enhance such collisions even in the case of a more slowly rotating  SMBH via  high particle orbital eccentricities generated by the eccentric Kozai-Lidov mechanism \cite{smadar2019}.  In summary, provided that black hole spins in galactic nuclei can reach near-extremal scales there may be a relatively low cost  astrophysical complement to construction of a FCC. 

{\it Acknowledgments –} This work was supported by a Leverhulme Trust International Professorship grant [number LIP-202-014]. For the purpose of Open Access, AM has applied a CC BY public copyright license to any Author Accepted Manuscript version arising from this submission. 

\bibliography{andy}
\end{document}